\documentclass[structabstract]{aa}

\usepackage{graphicx}
\usepackage{amsmath}
\usepackage{epsfig}
\usepackage{txfonts}

\begin{document}

   \title{Spectral modelling of massive binary systems}

   \author{M. Palate\inst{1}
          \and
          G. Rauw\inst{1}
          \and
          G. Koenigsberger\inst{2}
          \and
          E. Moreno\inst{3}
          }

   \institute{Institut d'Astrophysique et de G\'eophysique, 
   Universit\'e de Li\`ege, Bât. B5c, All\'ee du 6 Ao\^ut 17, 4000 Li\`ege, Belgium\\
             \email{palate@astro.ulg.ac.be}
             \and
             Instituto de Ciencias F\'{\i}sicas, Universidad Nacional Aut\'onoma de M\'exico, 
             Cuernavaca, Morelos 62210, M\'exico\\
							\email{gloria@astro.unam.mx}
             \and
             Instituto de Astronom\'{\i}a, Universidad Nacional Aut\'onoma de M\'exico, 
             M\'exico, D.F., M\'exico\\
             \email{edmundo@astro.unam.mx}
             }

   \date{Received $<$date$>$; accepted $<$date$>$}

	\abstract
   {The spectra of massive binaries may be affected by interactions between the stars in the system. These are believed to produce observational phenomena such as the Struve-Sahade effect.}
   {We simulate the spectra of massive binaries at different phases of the orbital cycle, accounting for the gravitational influence of the companion star on the shape and physical properties of the stellar surface.}
   {We used the Roche potential modified to account for radiation pressure to compute the stellar surface of close circular systems. We further more used the tidal interactions with dissipation of energy through shear code for surface computation of eccentric systems. In both cases, we accounted for gravity darkening and mutual heating generated by irradiation to compute the surface temperature. We then interpolated non-local thermodynamic equilibrium (NLTE) plane-parallel atmosphere model spectra in a grid to obtain the local spectrum at each surface point. We finally summed all contributions, accounting for the Doppler shift, limb-darkening, and visibility to obtain the total synthetic spectrum. We computed different orbital phases and different sets of physical and orbital parameters.}
   {Our models predict line strength variations through the orbital cycle, but fail to completely reproduce the Struve-Sahade effect. Including radiation pressure allows us to reproduce a surface temperature distribution that is consistent with observations of semi-detached binary systems.}
   {Radiation pressure effects on the stellar surface are weak in (over)contact binaries and well-detached systems but can become very significant in semi-detached systems. The classical von Zeipel theorem is sufficient for the spectral computation. Broad-band light curves derived from the spectral computation are different from those computed with a model in which the stellar surfaces are equipotentials of the Roche potential scaled by the instantaneous orbital separation. In many cases, the fit of two Gaussian/Lorentzian profiles fails to properly measure the equivalent width of the lines and leads to apparent variations that could explain some of the effects reported in the literature.}


   \maketitle

\section{Introduction}
Studing massive stars is important because they play a key role in galaxy evolution: their strong winds interact with the ambient interstellar medium and can trigger the formation of new stars, they produce heavy chemical elements, and they are an important source of UV radiation. Recent research indicates that a large part of massive stars ($\sim 50$\%) form a binary or a multiple system (Mahy et al. \cite{Mahy}, Sana \& Evans \cite{SanaEvans}). The spectra of binary systems are an invaluable source of information for determining the physical properties (such as masses, temperatures, and radii) of stars. It has been shown, however, that the observational analysis of massive binaries is complicated by effects that are linked to interactions in these systems that affect the spectra and the spectral classification (Sana et al. \cite{Sana}, Linder et al. \cite{Linder}, Linder \cite{LinderPhDTh}, and references therein). Unfortunately, the majority of spectral modelling codes are designed for isolated spherical stars and cannot reproduce the particular effects of the binarity. Therefore, it is important to improve our models of spectral computation to accurately represent the spectra of these stars. In this context, we have developed a method that takes into account some of the effects that are produced by gravitational interactions in the system. 

The first models of binary systems in which the stars are not spherical have been proposed by Russell \& Merrill (\cite{Russell}) to reproduce the light curves of binaries. Then, Kopal (\cite{Kopal}) introduced the Roche potential approach that was first used by Lucy (\cite{Lucy}) and Wilson \& Devinney (\cite{WilsonDevinney}) followed by numerous other works. In a first step, we designed a similar code for circular massive binary systems (Palate \& Rauw \cite{Palate}, hereafter Paper I) in which the distorted shapes of the stellar surfaces are caused by gravitational interaction. The emerging spectra are then computed accounting for the shape of the stars and their local properties. In this paper we extend the method to incorporate radiation pressure effects on the stellar surface and the case of  eccentric binaries. This new version of our CoMBiSpeC (code of massive binary spectral computation) model now allows spectral computation of (almost) any massive binary system. 

\section{Improvements of the method}\label{sec:2}
	Our method for modelling the spectra of circular binary systems, presented in Paper I, consists of computing the actual stellar surface following the Roche potential approach and accounting for the local gravity and temperature on each surface element. Non-LTE spectral model grids (TLUSTY, Lanz \& Hubeny \cite{Lanza}, \cite{Lanzb}) are used to compute the integrated spectrum of the star at each orbital phase (see also, e.g., Linnell \& Hubeny \cite{Linnell(1)} and Linnell et al. \cite{Linnell(2)}). Because our spectra are synthetic, they are free of observational noise. In this section we describe the two improvements that have been implemented to the method. The first is the inclusion of radiation pressure effects; i.e., the irradiation of each star by the companion and the effect of the radiation on each star's own surface. The second is the use of an explicit calculation of the stellar surface, from first principles, which here is applied to the case of eccentric binaries. In eccentric systems, the orbital separation changes as a function of phase, and accordingly, the deformations of the surface and the orientation of the tidal bulge are also variable.
	
	\subsection{Radiation pressure effects}
		There exists quite an extensive body of literature (see e.g.  Dermine et al. \cite{Dermine}, Howarth \cite{Howartha}, Schuerman \cite{Schuerman}, Drechsel et al. \cite{Drechsel} and Phillips \& Podsiadlowski \cite{Phillips}) on the impact of radiation pressure on the Roche potential in early-type binaries. According to Dermine et al. (\cite{Dermine}), the effects of radiation pressure in a binary system can be separated into three parts: radiation pressure on each star's own surface, the effect on the companion, and radiation pressure on the matter outside the binary orbit. Here we focus on the two first effects, which are called internal and external radiation pressure (Drechsel et al. \cite{Drechsel}). The impact of the radiation pressure on the matter outside the binary is, of course, very important for the surrounding medium but is irrelevant for calculating the shape of the stars.
	 Howarth (\cite{Howartha}) pointed out that the effect of the internal radiation pressure can be treated as a simple scaling of the Roche potential of the stars. Another approach, proposed first by Schuerman (\cite{Schuerman}), consists of scaling the mass of the stars rather than the entire potential. The former approach is more appropriate because the emitted radiation bolometric flux and hence the radiation pressure scale with the local gradient of the potential (which also includes the attraction by the other star and the centrifugal forces). Therefore, according to Howarth (\cite{Howartha}), the stellar topology is not affected by internal radiation pressure and the latter simply scales the local surfaces gravity. The Newtonian gravity is defined by
		
		\begin{equation}
				\vec{g} = - \vec{\nabla} \phi,
		\label{eq:gravpotscal}
		\end{equation}
		where $\phi$ is a scalar gravity potential per unit mass.
	
		The von Zeipel (\cite{VonZeipel}) theorem can be written\footnote{Throughout this paper, the temperature is to be understood as effective temperature.}
		\begin{equation}
				T^4 \propto \left\Vert \vec{g} \right\Vert.
		\end{equation}
		
		The magnitude of the radiative acceleration is given by (Howarth \cite{Howartha})
		\begin{equation}
			  a_{rad}  =  - \frac{\kappa\pi}{c}\mathcal{F},
		\end{equation}	  
		where $\kappa$ is the flux mean opacity per unit mass and $\mathcal{F}$ is the astrophysical bolometric flux ($\mathcal{F}\propto T^4$).
		
		If we define $\Gamma$ as the ratio of the radiative to the gravitational acceleration, $a_{rad}=\Gamma g$. According to Howarth (\cite{Howartha}), this ratio is constant for a given star and independent of the position on the stellar surface since both accelerations vary with $T^4$. Therefore, we have that the effective potential is the potential without radiation pressure minus the radiative acceleration
		\begin{equation}
	  \vec{g}_{eff} = (1-\Gamma) \vec{g}.
		\end{equation}
	
		From equation \ref{eq:gravpotscal}, $\phi_{eff} = (1-\Gamma)\phi$ and the ratio $\Gamma$ can be written as (following the Castor et al. \cite{CAK} theory for evaluating $\kappa$)
		
		\begin{equation}
				 \begin{array}{r c l}
				 \Gamma & = & \dfrac{\kappa\pi}{c}\mathcal{F}/g\\[3mm]
				 				& = & \dfrac{\sigma_{Th}}{m_{H}c}\sigma T^4/g,\\
		 \end{array}
		\end{equation}
		with $\frac{\sigma_{Th}}{m_{H}} \approx 0.036$ m$^2$kg$^{-1}$, $\sigma_{Th}$ the Thomson-scattering cross section and $\sigma$ the Stefan-Boltzmann constant.
		
	Because $\Gamma$ is independent of the position on the stellar surface, we can compute it for $T = T_{pole}$ and $g = g_{pole}$ 
		\begin{equation}
			\Gamma = \frac{\sigma_{Th}}{m_{H}c}\sigma T^4_{pole}\frac{1}{\left\Vert \vec{g}_{pole} \right\Vert}.
		\end{equation}
		
	 The advantage of this method is that the von Zeipel theorem remains valid. Another advantage is the simplicity of the method which does not require many computational resources.
	 
	 The external radiation pressure is more difficult to treat. We have based our method on the approach of Drechsel et al. (\cite{Drechsel}) and Phillips \& Podsiadlowski (\cite{Phillips}). This method consists of scaling the mass of the companion in the Roche potential. The scale parameter $\delta = \frac{F_{rad}}{F_{grav}}$ is computed iteratively for each surface point. The external radiation pressure can be seen as a force that decreases the attraction of the companion. Therefore, the companion seems less massive, and consequently the mass has to be scaled. The new ``Roche'' potential can be written
		
		\begin{equation}
		 \begin{array}{r c l}
				\Omega & = & \dfrac{1}{r}+\dfrac{q(1-\delta(r, \theta, \varphi))} {\sqrt{r^{2}-2r\cos\varphi\sin\theta+1}}+\dfrac{q+1}{2}\cdot r^{2}\sin^{2}\theta\\
						 		&& \multicolumn{1}{r}{-qr\cos\varphi\sin\theta},\\
		 \end{array}
		\end{equation}				
		where $r=\sqrt{x^{2}+y^{2}+z^{2}}$, $q=\frac{m_{2}}{m_{1}}$, $x=r\cos\varphi\sin\theta$, $y=r\sin\varphi\sin\theta$, and $z=r\cos\theta$.
		\noindent Here, $\theta$ and $\varphi$ are the colatitude and longitude angle in the spherical coordinates centred on the star. $\theta=0$ and $\varphi=0$ correspond respectively to the north pole of the star and the direction towards its companion.
	
	The radiation pressure can be written
		\begin{equation}
			  P_{rad}  =  \frac{1}{c}\int_0^{\infty}\int_{\omega} I_\nu\,\cos^2\Psi_1 \,\mathrm{d}\omega \mathrm{d}\nu,
		\end{equation}
	\noindent where $\Psi_1$ is the angle between a local surface normal on the irradiated star and the direction to a surface element on the other star. $ I_\nu$ is the specific intensity of the external radiation field, and $\nu$ is the frequency.
	 
	The radiation pressure gradient per unit mass is equal to (in plane-parallel approximation)
		\begin{equation} \label{eq:Frad}
				 \begin{array}{r c l}
				 F_{rad} & = & -\dfrac{1}{\rho} \dfrac{\mathrm{d}P_{rad}}{\mathrm{d}r}\\
				 				& = & \frac{1}{c}\int_{\omega}\int_0^{\infty} \kappa_{\nu} \dfrac{\mathrm{d}I_{\nu}} {\mathrm{d}\tau_\nu} \cos^2\Psi_1 \,\mathrm{d}\nu \mathrm{d}\omega \\
									& = & \frac{1}{c}\int_{\omega}\int_0^{\infty} \kappa_{\nu} I_{\nu}\,\cos\Psi_1 \,\mathrm{d}\nu \mathrm{d}\omega. \\
		 \end{array}
		\end{equation}

		Following Castor et al. (\cite{CAK}), we assume that the dominant source of opacity comes from electron scattering, thus, we have $\kappa_{\nu} = \frac{\sigma_{Th}}{m_H}$.
		
		The value of $\delta$ is therefore computed with an iterative method. We start from the classical unmodified Roche potential and compute the radiation received by each surface element from its companion. The radiation is computed by summing the contribution of all points that are visible from the primary (resp. secondary) on the secondary (resp. primary). The method applied to compute the received bolometric flux by a point is derived from equation \ref{eq:Frad} and is similar to the reflection effect treatment of Wilson (\cite{Wilson}). Indeed, we have $I_{\nu} = I_{\nu,0}\times(1-u+u\cos\Psi_2)$, where $\Psi_2$ is the angle between a local surface normal on the emitting star and the direction towards an irradiated surface element on the other star, and $u$ is the linear limb-darkening coefficient based on the tabulation of Claret \& Bloemen (\cite{Claret}). Moreover, we can write $\mathrm{d}\omega = \frac{\mathrm{d}S_2 cos\Psi_2}{s^2}$, where $s$ is the distance between the irradiated and emitting surface. Finally, if the radiation comes only from the companion, we have $\int_0^{\infty} I_{\nu,0}\,\mathrm{d}\nu = \frac{\sigma T^4}{\pi}$. The local $\delta$ factor is thus defined by $\delta = \frac{F_{rad}*d^2}{GM}$, where $d$ is the distance between the point and the centre of the emitting star. This factor can be written (for the radiation emitted by the secondary and received by the primary) as
	\begin{equation}
		  \delta_{2}  =  \frac{\sigma\sigma_{th}d^2}{\pi c m_HGM_2}\sum\limits_{\substack{cos\Psi_1>0 \\ cos\Psi_2>0}}^{}{T_2^4 (1-u+u\cos\Psi_2) \cos\Psi_1\cos\Psi_2 \frac{\mathrm{d}S_2}{s^2}}.
		\end{equation}
		Then, with the $\delta$ factor, we recompute the stellar surface with the modified Roche potential. Again, we evaluate the $\delta$ factor and surface until we reach convergence.	
	
	\subsection{Eccentricity} \label{subsec:Eccentricity}
		The extension to eccentric systems uses the TIDES\footnote{Tidal interactions with dissipation of energy through shear.} code (Moreno et al. \cite{Morenoa}, \cite{Morenob}, \cite{Morenoc}). This code computes the time-dependant shape of the stellar surface for eccentric and/or asynchronous systems. It also provides velocity corrections for rigid body rotation. The method consists of modelling a deformable and perturbed surface layer that lies upon a rigid body that is in uniform rotation by solving the equations of motion of a grid of surface elements. These equations take into account centrifugal and Coriolis forces, gas pressure, and viscous effects in addition to those of the gravitational potential of both stars. It is important to note that in asynchronously rotating or eccentric systems, the linear approximation of the Roche potential is no longer valid and the presence of viscosity in the medium may lead to non-linear effects\footnote{The viscous stresses are included in the equations of motion and connect the surface elements  to each other and to the inner rigid body.}. The surface shape and local velocities are obtained as a function of orbital phase from the solution of the equations of motion. Full details of the model are given in Moreno \& Koenigsberger (\cite{Morenoa}), Toledano et al. (\cite{Toledano}), and Moreno et al. (\cite{Morenob}, \cite{Morenoc}).
		
	With the derived values of radius and velocity perturbations, we compute the surface gravity and temperature. Assuming the perturbations to be small, the gravity is computed by using a classical Roche gradient.
	
	 The temperature distribution is computed following the von Zeipel theorem
		\begin{equation}
		T_{\mathrm{local}} = T_{\mathrm{pole}}\left(\dfrac{\left\Vert\underline{\nabla}\Omega_{\mathrm{local}}\right\Vert} {\left\Vert\underline{\nabla}\Omega_{\mathrm{pole}}\right\Vert }\right)^{0.25p},			
		\end{equation}
\noindent where the gravity-darkening parameter $p=1$ in the case of massive stars.
		
		We also include reflection effects that heat the surface of the star through irradiation of each component by the other. The effect was treated by following the approach of Wilson (\cite{Wilson}). We neglected any cross-talk between the surface elements (e.g. due to horizontal advection or radiative exchanges between neighbouring surface elements of different temperature). This is probably an important approximation because this effect might smooth out some of the temperature variations, especially near periastron (for eccentric systems) where the stars move faster and where the gravitational interactions are stronger. These interactions and velocity imply that energy dissipation due to viscosity effects is more significant near periastron and thus the cross-talk is expected to be higher. 
				
		The spectral computation was performed with the CoMBiSpeC code, which was described in Paper I. For this paper, some minor changes have been applied to take into account particularities linked to the eccentricity, such as the variation of the separation between the two stars.
		
\section{Circular-orbit models} \label{sec:3}

	We present here models of four binary systems with circular orbits. Three of them were previously studied in Paper I, i.e., models 1, 2, and 3. These models are based on the HD\,159176, HD\,165052, and HD\,100213 systems analysed by Linder et al. (\cite{Linder}). The fourth system, model 4 in this paper, is based on Sk-67\degr105 studied by Bonanos (\cite{Bonanos}). All four of these systems are in synchronous rotation and, thus, are in equilibrium configuration so that the Roche potential approach is valid. For all models, we compared the shape, temperature, and gravity variations caused by radiation pressure. Table \ref{tab:PARAMcirc} gives a summary of the parameters (from  Linder et al. \cite{Linder} and Bonanos \cite{Bonanos}) used for the computation. We have chosen the polar temperature (rather than a surface mean temperature) as the reference effective temperature in Table \ref{tab:PARAMcirc}. This is because $T_{pole}$ is used to compute the local temperature via the von Zeipel theorem and remains essentially unaffected by radiation pressure effects. Finally, for model 4, we investigated the Struve-Sahade effect for some lines.

	\begin{table*}[ht!]
		\caption{Parameters (from  Linder et al. \cite{Linder}: models 1 to 3, and Bonanos \cite{Bonanos}: model 4) of the circular binary systems simulated in this paper. Inclinations in brackets stand for non-eclipsing systems.}
		\label{tab:PARAMcirc}
		\centering
			  \begin{tabular}{l c c c c c}
				\hline\hline 
				Parameters & Model 1 & Model 2 & Model 3 & Model 4\\
				\hline
				Period (day) & $3.36673$ & $2.95515$ & $1.3872$ & $4.251$\\
				Mass ratio & $0.96$ & $0.87$ & $0.68$ & $0.42$\\
				Semi-major axis ($R_{\sun}$) & $38.23$ & $31.25$ & $17.34$ & $38.94$\\
				Inclination (\degr) & $(48)$ & $(23)$ & $77.8$ & $89.9$\\
				Mass of primary ($M_{\sun}$) & $33.8$ & $25.15$ & $21.7$ & $30.9$\\
				Mass of secondary ($M_{\sun}$) & $32.41$ & $21.79$ & $14.7$ & $13.0$\\
				Primary polar temperature (K) & $38000$ & $35500$ & $35100$ & $35000$\\
				Secondary polar temperature (K) & $38000$ & $34400$ & $31500$ & $32500$\\
				Polar radius of primary ($R_{\sun}$) & $9.37$ & $9.10$ & $6.74$ & $15.1$\\
				Polar radius of secondary ($R_{\sun}$) & $8.94$ & $8.47$ & $5.62$ & $11.14$\\
				\hline
			\end{tabular}
	\end{table*}

	\subsection{Models 1 and 2}
\begin{figure}[ht!]
			\resizebox{\hsize}{!}{\includegraphics{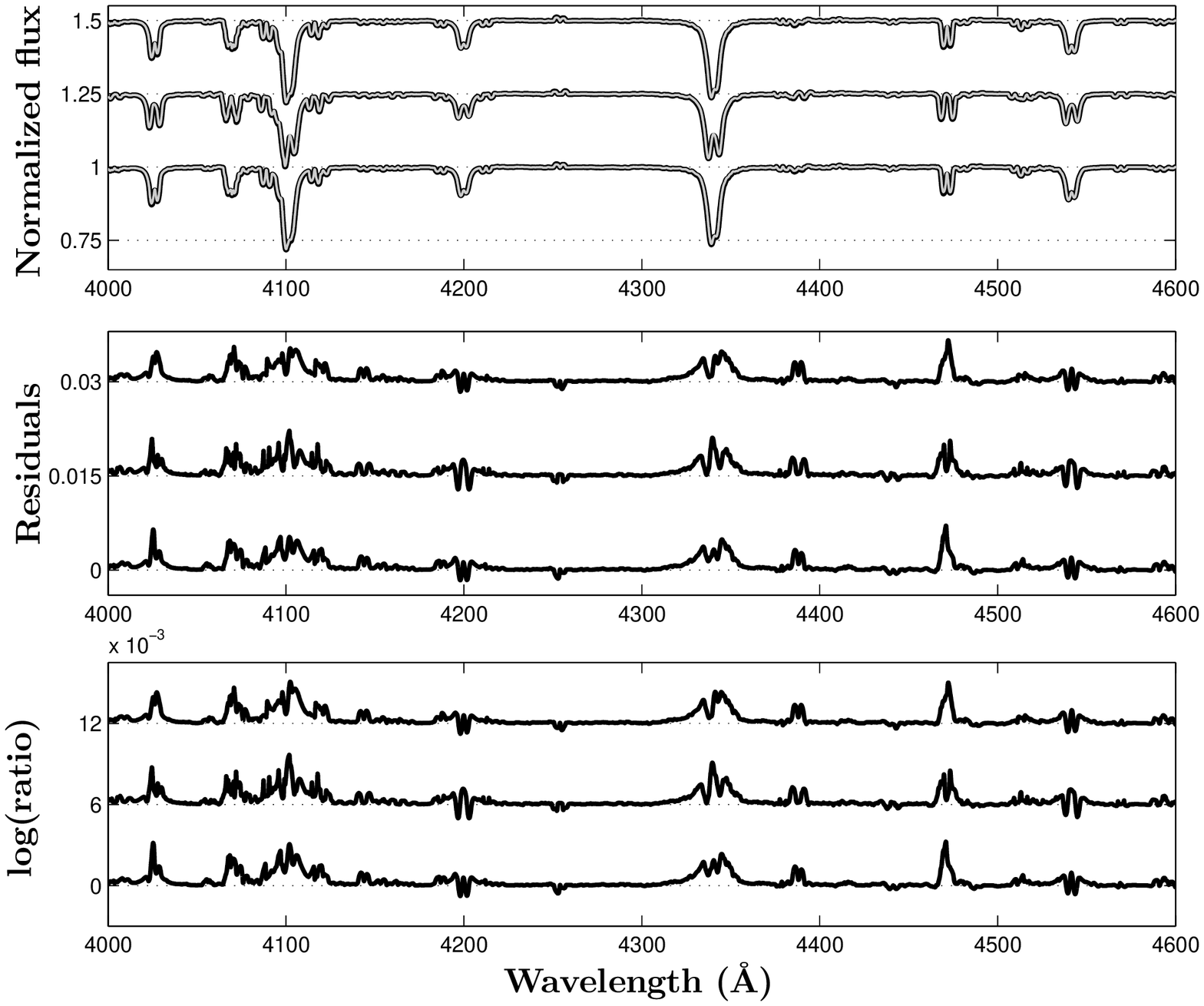}}
			\caption{\textit{Top}: Comparison between the normalized spectra computed without radiation pressure (\textit{in black}) and with radiation pressure (\textit{in grey}) for model 1. To better distinguish spectra, the black line is wider than the grey line. Orbital phases are, from bottom to top: phases $0.1$, $0.25$ and $0.4$. The spectra are shifted vertically by $0.25$ continuum units for clarity. \textit{Middle}: Corresponding residuals (spectrum with radiation pressure minus spectrum without radiation pressure). The residual plots are shifted by $0.015$ for clarity. \textit{Bottom}: Logarithm of the ratio of the fluxed spectra with radiation pressure divided by the one without. These plot are shifted by $0.006$ for clarity.}
			\label{fig:Spectramodel1}
		\end{figure}

	We have analysed the spectra\footnote{Throughout this paper, the word ``spectrum'' refers to a synthetic spectrum.} of models 1 and 2 at 20 orbital phases and found that the impact of radiation pressure is rather low. The comparison was made by subtracting the normalized spectra with radiation pressure to the spectra without radiation pressure. On each of these ``residual'' spectra, we searched for the maximum, mean, and median values and the standard deviation. Then we searched for the maximum of the maxima (over the orbital phase) and computed a mean of the medians, the standard deviations, and means. We also looked for the spectral bins that present deviations higher than $1$\% (over a total of 360000 bins). Table \ref{tab:comp prad} gives a summary of these values for the combined spectra of the studied models. For example, the highest differences are $0.0113$ and $0.0191$ for models 1 and 2. These maxima occur in both cases at phase $0.5$, i.e. at the conjunction with the primary in front of the secondary. These differences are significant but, the mean\footnote{We only took into account the spectral bins that include spectral lines for this mean value.} difference is only $\sim$$7-9\times 10^{-4}$, which is very small. The general appearance of the spectra is preserved and the conclusions reached in Paper I remain valid. We can see in Fig. \ref{fig:Spectramodel1} that the spectra of model 1 with and without radiation pressure are nearly indistinguishable (\textit{top panels}).
	
	The largest and mean radii of the stars are smaller if radiation pressure effects are included.	The radiation pressure also decreases the mean $\log(g)$ by $\sim 0.05$ despite the smaller radii. Finally, the mean temperatures increase by $100$K. This increase of the temperature and decrease of gravity cause a weak increase of a few per cent of the total flux emitted by the stars. Fig. \ref{fig:Spectramodel1}, \textit{lower panel}, displays the $\log$ of the ratio of the fluxed spectra with and without radiation pressure and underlines the small differences in the continuum. Table \ref{tab:TR} gives a summary of the radii and temperatures of circular models.
	
	\begin{table*}[ht]
		\caption{Comparison between the spectra with and without radiation pressure.}
		\label{tab:comp prad}
		\centering
			\begin{tabular}{l c c c c c} 
				\hline\hline 
				Models & $\left\Vert\Delta_{max}\right\Vert$ & $\left\Vert\Delta_{mean}\right\Vert$ & $\left\Vert\Delta_{median}\right\Vert$ & $\left\Vert\sigma_{\Delta}\right\Vert$ & $\Delta>1$\%\\
				\hline
				Model 1 & $0.0113$ & $6.8\times 10^{-4}$ & $3.2\times 10^{-4}$ & $0.0014$ & $12$\\
				Model 2 & $0.0191$ & $8.7\times 10^{-4}$ & $3.8\times 10^{-4}$ & $0.0018$ & $213$\\
				Model 3 & $0.0063$ & $5.1\times 10^{-4}$ & $3.0\times 10^{-4}$ & $8.7\times 10^{-4}$ & $0$\\
				Model 4 & $0.1247$ & $0.0069$ & $0.0022$ & $0.0171$ & $95820$\\
				\hline
			\end{tabular}
	\end{table*}
	
	\begin{table}[ht]
		\caption{Temperature and radius of stars at different locations at the stellar surface.}
		\label{tab:TR}
		\centering
			\begin{tabular}{l c c c c c c}
				\hline\hline
				Stars & $R_{\text{side}}$ & $R_{\text{point}}$ & $R_{\text{back}}$ & $T_{\text{side}}$ & $T_{\text{point}}$ & $T_{\text{back}}$\\
				 & ($R_{\sun}$) & ($R_{\sun}$) & ($R_{\sun}$)& (K) & (K) & (K) \\
				\hline	
				Model 1$^{\text{1}}$ & $9.51$ & $9.46$ & $9.70$ & $37442$ & $38061$ & $36705$\\
				Model 1$^{\text{2}}$ & $9.06$ & $8.97$ & $9.23$ & $37496$ & $38271$ & $36800$\\
				Model 2$^{\text{1}}$ & $9.32$ & $9.44$ & $9.62$ & $34654$ & $34937$ & $33577$\\
				Model 2$^{\text{2}}$ & $8.66$ & $8.67$ & $8.96$ & $33643$ & $34465$ & $32510$\\
				Model 3$^{\text{1}}$ & $7.12$ & $8.44$ & $7.63$ & $33198$ & $29306$ & $29211$\\
				Model 3$^{\text{2}}$ & $5.87$ & $6.92$ & $6.43$ & $30112$ & $30855$ & $27369$\\
				Model 4$^{\text{1}}$ & $15.80$ & $15.95$ & $16.42$ & $33411$ & $33882$ & $32121$\\
				Model 4$^{\text{2}}$ & $11.61$ & $10.72$ & $12.89$ & $31185$ & $34062$ & $27912$\\
		  	\hline
			\end{tabular}
			\tablefoot{$^{\text{1}}$: primary star, $^{\text{2}}$: secondary star. \textit{side}: $r(\phi,\theta) = r(\pi/2,\pi/2)$, \textit{point}: $r(\phi,\theta) = r(0,\pi/2)$, \textit{back}: $r(\phi,\theta) = r(\pi,\pi/2)$}
	\end{table}
		
	\subsection{Model 3}
\begin{figure}[ht!]
				\resizebox{\hsize}{6cm}{\includegraphics{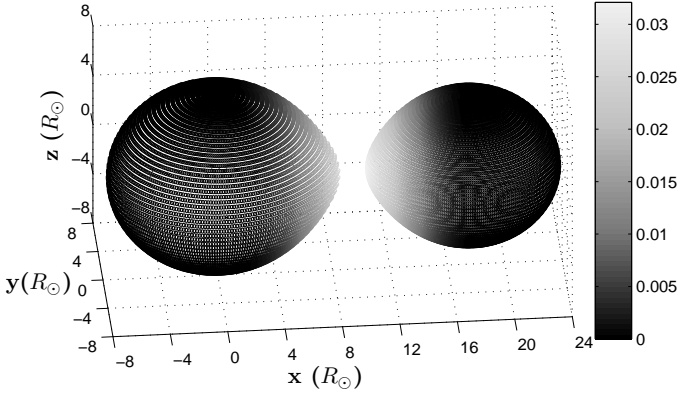}}
			\caption{Value of the radiation pressure parameter, $\delta$, over the stellar surface for model 3. The highest values are located near the L1 point.}
			\label{fig:Model3delta}
		\end{figure}

	This model leads to an (over)contact binary if radiation pressure is not included, but when it is included, the system becomes detached \footnote{Strictly speaking, the CoMBiSpeC model cannot handle overcontact configuration, but the most extreme cases correspond to situations where both stars fill their Roche lobe.}. Surprisingly, however, the impact of radiation pressure on the spectrum is weak. This is because the parts of the star that contribute most to the spectrum come from the rear and side parts of the star which are less modified by the radiation pressure because $\delta$ is small (see Fig.\ref{fig:Model3delta}). Moreover, the shape of the star is not strongly affected by the radiation of the companion. This is {\it a priori} unexpected but the reason for it is that the most deformed parts of the stars mainly see the coldest part of the companion, which leads to a rather low value of the maximum $\delta$. We stress here that in systems where the two stars are in (over)contact (or nearly so) and are of similar size and temperature, the reflection effect is not sufficient to counterbalance the gravity darkening. Thus, the coldest parts of the stars are those near the L1 point.
	
	In Paper I, we pointed out that we failed to reproduce the surface temperature distribution deduced by Linder et al. (\cite{Linder}) for HD\,100213. These authors showed in their analysis through radial velocity measurements that the He\,II and He\,I lines were not formed in the same region of the stellar surfaces. The small changes of the stellar surface induced by the radiation pressure increase the importance of the reflection process and are sufficient to now explain the surface temperature distribution observed by Linder et al. The He II lines have a smaller radial velocity amplitude and thus are apparently stronger in the hemisphere facing the companion, whilst the He\,I lines have larger velocity amplitudes, indicating that they are stronger on the opposite hemisphere. This suggests that the hemisphere facing the companion is hotter than the opposite one. When radiation pressure is added, the stars are less deformed. This implies a weak temperature increase of the facing hemispheres caused by the decreasing gravity darkening and a higher effectiveness of the reflection process. This weak increase is therefore sufficient to explain the observations.
		
 Espinosa Lara \& Rieutord (\cite{Espinosa}) and Maeder (\cite{Maeder}) have suggested that von Zeipel's (\cite{VonZeipel}) classical theorem requires modifications. These theoretical studies present more complex models that could be, at first approximation, equivalent to the classical von Zeipel theorem with a gravity-darkening parameter (GDP) $p$ smaller than $1.00$. Therefore, we chose this very deformed system in which the von Zeipel theorem implies strong temperature variations to test the influence of $p$ on the spectra. We computed the spectrum for values of the GDP of $0.25$, $0.50$, $0.75$, $1.25$ and $1.50$, all other parameters having the same value as model 3 in Table \ref{tab:PARAMcirc}. We found that for a moderate change of $p$ (of $+/-0.25$) the spectra are not notably affected. We also found that the variations remain within the noise level of real data for variations of up to $+/-0.50$. Therefore, since we made some other important assumptions (no cross-talk, assumptions underlying the Roche potential), it is not useful to adopt another more complex formalism for gravity darkening because it does not notably affect our simulations.
	
	\subsection{Model 4}
\begin{figure}[ht!]
				\resizebox{\hsize}{12cm}{\includegraphics{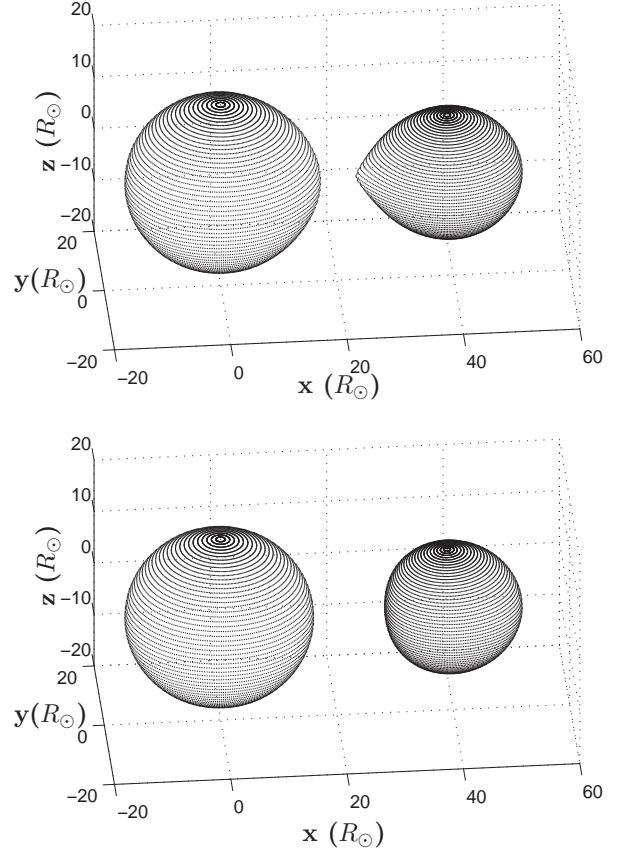}}
			\caption{Surface modification by the radiation pressure for model 4. \textit{Top:} Star surface without radiation pressure effect. \textit{Bottom:} Star surface with radiation pressure effect.}
			\label{fig:Model4surf}
		\end{figure}

This system was chosen as an example of one in which the radiation pressure considerably modifies the shape of the stars (see Fig. \ref{fig:Model4surf}), underlining the potential significance of this effect. The system parameters (see Table \ref{tab:PARAMcirc}) are based on the observational study of Sk-67\degr105 that was carried out by Bonanos (\cite{Bonanos}).

		\subsubsection{Spectral classification}
		Observational studies have shown that the primary and secondary are an O7-8V and an O8-8.5 III-V star.	We first determined the spectral type of our synthetic spectra with the Conti (Conti \& Alschuler \cite{Conti}) -- Mathys (\cite{Mathys(1)}, \cite{Mathys(2)}) criterion (hereafter Conti--Mathys criterion). This criterion is based on the ratio of the equivalent widths (EWs) of He\,I $\lambda$\,4471 to He\,II $\lambda$\,4542 for the spectral-type determination and on the ratio of the EWs of Si\,IV $\lambda$\,4089 to He\,I $\lambda$\,4143 for the luminosity-class determination. We found that our model spectra that include radiation pressure correspond to an O6.5-7I type for the primary spectrum and an O7.5-O8I type for the secondary. The determination of the spectral type gives the same result whether the measurement is made on individual spectra or on combined spectra\footnote{``Combined spectrum'' refers to the spectrum of the primary plus the spectrum of the secondary, i.e. the spectrum of the entire system.}. In our model, the stars seem hotter, and the problem encountered in Paper I concerning the luminosity class is again present because we overestimate the luminosity. In a second approach, we used the Walborn \& Fitzpatrick (\cite{Walborn}) atlas to compare individual spectra to the reference atlas and thus re-classify the stars. With the atlas we found for the primary and secondary an O(6.5)-7 V and an O8-8.5 (III)-V classification. The classification using the atlas is of course more qualitative, therefore we give a range and the best correspondence is in brackets. Within the uncertainties, the agreement with the observations is good.
		
		\subsubsection{Struve-Sahade effect}
		The Struve-Sahade effect (hereafter S-S effect) was originally defined as the apparent strengthening of the secondary spectrum when the star is approaching the observer and its weakening as it moves away (Howarth et al. \cite{Howarthb}, Linder et al. \cite{Linder}). A more general definition is the apparent variation of the line strengths of either of the binary components as a function of the orbital phase (Howarth et al. \cite{Howarthb}). Bonanos (\cite{Bonanos}) reported a quite considerable S-S effect for several spectral lines of Sk-67\degr105. We studied eight of these lines in detail: He\,I $\lambda\lambda$\,4026, 4143, 4471, 4713, 5016, Si\,IV $\lambda$\,4089, He\,II $\lambda\lambda$\,4542, and 5411.
		
		First, we analysed the EWs on individual spectra of the primary and the secondary\footnote{The mean relative error of the normalized EWs can be estimated to $\sigma = ~10^{-3}-10^{-2}$. The $\sigma$ given by the deblend/line routine is of the same order of magnitude. This error estimation is valid for all studied lines throughout this paper.}. The measurements of the line EWs were performed with the MIDAS software developed by ESO. The EWs of lines in the spectra of the individual stars were determined directly by simple integration with the integrate/line MIDAS routine, whilst for the binary system, we used the deblend/line command as we would do for actual observations of a real binary spectrum. The latter routine fits two Gaussian line profiles to the blend of the primary and secondary lines. 
		
\begin{figure}[ht!]
		  \resizebox{\hsize}{4cm}{\includegraphics{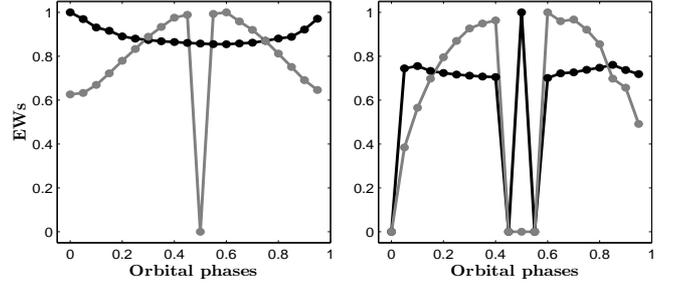}}
			\caption{Example of EW variations of the He\,II $\lambda$\,4542 line. The EWs have been normalized to the highest value. \textit{Black:} Primary. \textit{Grey:} Secondary. \textit{Left}: Variations measured on individual spectra (EW$_{\text{p, max}}=0.53$\AA, EW$_{\text{s, max}}=0.47$\AA). \textit{Right}: Variations measured on combined spectra by fitting two Gaussians to the blended lines (EW$_{\text{p, max}}=0.52$\AA, EW$_{\text{s, max}}=0.16$\AA).}
			\label{fig:EWmodel4}
		\end{figure}

		The EWs of the different lines measured on the individual spectra display phase-locked variations. For the primary, the EWs of the He\,I lines reach a maximum at phase $0.5$ (when the primary eclipses the secondary) because the coldest part of the star is the rear part. Two minima are visible at phases $0.05-0.1$ and $0.9-0.95$ because the observer starts to see the front part of the primary star. This part of the star is the hottest because of the very effective reflection in this system. The minimum is not observed at phase $0$ because of the roughly annular eclipse that hides a substantial part of the stellar surface. The variation amplitude is line-dependent with $\sim 5$\% for He\,I $\lambda$\,4026, $60$\% for He\,I $\lambda$\,4143, and $20$\% for the other He\,I lines. As expected for heating effects, the He\,II lines exhibit a reverse variation compared to the He\,I lines (see Fig.\ref{fig:EWmodel4}). The Si\,IV line displays a similar variation as the He\,I lines. 
		
	For the secondary, the EWs display the same variation as the primary, but shifted by $0.5$ in phase. We can notice the zero value at phase $0.5$ when the secondary is totally eclipsed. The variation amplitude of the He\,I $\lambda\lambda$\,4026, 4713, 5016, and Si\,IV lines is small and less than $\sim 10$\%. The amplitude is greater than $\sim 20$\% for the other lines (we did not account for the zero value).
		
		The measurements of the combined spectra using a two-Gaussian fit to the blended lines agree relatively well with the variations measured on individual spectra for several lines: He\,I $\lambda\lambda$\,4026, 4143, 4713, 5016, He\,II $\lambda\lambda$\,4542, and 5411. The variation amplitude is sometimes overestimated, though. The EWs of He\,I $\lambda$\,4471 are asymmetric before and after phase $0.5$. The variation does not seem regular but is similar to the variations measured on individual spectra, i.e. a decrease of the EW of the secondary and an increase for the primary during the first half of the orbital cycle. Finally, the Si\,IV line displays irregular variations.
		
		Bonanos (\cite{Bonanos}) reported a strong S-S effect in Sk-67\degr105 for several lines that we studied in the model. The variations observed are consistent with the definition of Linder et al. (\cite{Linder}), i.e. in the first half of the orbital cycle the secondary (resp. primary) line is deeper than the primary (resp. secondary) line and the situation is reversed in the second half of the orbital cycle. However, even though some lines in our synthetic spectra display strong EW variations, we did not observe this type of phase dependence.
		
		\subsubsection{Radial velocity and synthetic broad-band light curve}
		The last two characteristics that we studied are the light curve in the wavelength range $3800-7100$ \AA\ and the radial velocity curve. The two eclipses are clearly visible in the light curve. The depth of the eclipses agrees well with the observations. We stress that light curves are ``by-products'' of the spectral computation and that CoMBiSpeC is not primarily designed for light curve computation. The semi-amplitude of the radial velocity also agrees well with the observations. The radial velocity was computed by taking the mean value of the velocity of the visible points of the stellar surface at a given phase weighted by the surface projected along the line of sight. The values are given for the primary and the secondary (observational value in brackets):\,$138.3$ ($137$) km\,s$^{-1}$ and $332.9$ ($326$) km\,s$^{-1}$. Finally, the synthetic light curve and radial velocity curve agree well with those derived by Bonanos (\cite{Bonanos}).
		
\section{Eccentric models } \label{sec:4}
	We present here five models of eccentric binary systems. These models are inspired by the following massive binaries: HD\,93205, HD\,93403, HD\,101131, HD\,152218, and HD\,152248, and are named models E1, E2, E3, E4, and E5. The stars were studied by: Antokhina et al. (\cite{Antokhina}), Rauw et al. (\cite{Rauw}), Gies et al. (\cite{Gies}), Sana et al. (\cite{Sanab}), Mayer et al. (\cite{Mayer}), and Sana et al. (\cite{Sanaa}). The parameters used for the computation are given in Table \ref{tab:PARAM}. We studied the spectral classification, the radial velocity curves, the broad-band light curve, and finally the S-S effect. We computed the surface of these stars with the TIDES code (Table \ref{tab:PARAM} \textit{bottom} gives the specific parameters used in the TIDES code) and used a modified version of the CoMBiSpeC code to compute the gravity and temperature. These quantities are not uniformly distributed across the stellar surface and, in addition, change over the orbital cycle. 
	
\begin{figure}[ht]
		  \resizebox{\hsize}{7cm}{\includegraphics{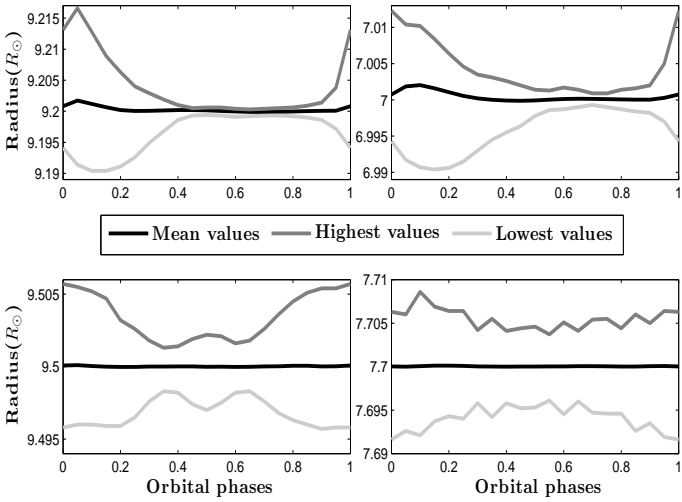}}
			\caption{Variations of the mean, highest and lowest visible radius (\textit{left}: the primary radii, \textit{right}: the secondary radii). \textit{Top}: Variations for model E1. \textit{Bottom}: Variations for model E3.}
			\label{fig:radiusE1E3}
		\end{figure}

\begin{figure}[ht]
		  \resizebox{\hsize}{8cm}{\includegraphics{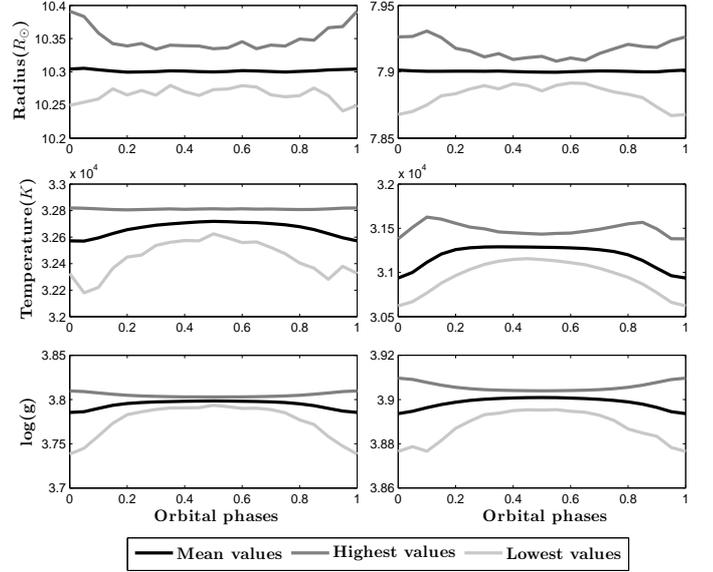}}
			\caption{Variations of the mean, highest and lowest visible radius, temperature, and log(g) for model E4. \textit{Left}: Values for the primary star. \textit{Right}: Values for the secondary star. }
			\label{fig:radiusE4}
		\end{figure}

\begin{figure}[ht]
		  \resizebox{\hsize}{7cm}{\includegraphics{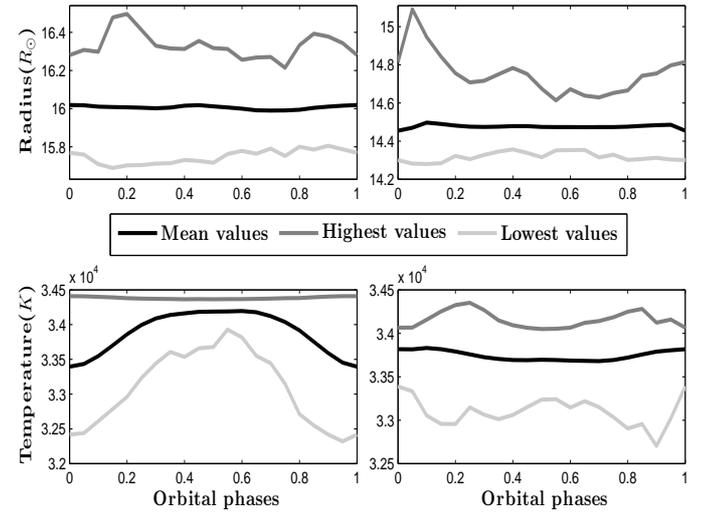}}
			\caption{Variations of the mean, highest and lowest visible radius and temperature for model E5. \textit{Left}: Values for the primary star. \textit{Right}: Values for the secondary star. }
			\label{fig:radiusE5}
		\end{figure}

	A sample of the phase-dependent behaviour of the mean radius, gravity, and temperature as well as their maximum and minimum values are shown in Figures \ref{fig:radiusE1E3}, \ref{fig:radiusE4}, and \ref{fig:radiusE5}. The maximum radius is largest near periastron (just before and/or just after). A model based on an instantaneous Roche lobe, such as the \textit{Nightfall} \footnote{For details see the Nightfall User Manual by Wichmann (1998) available at the URL: http://www.hs.uni-hamburg.de/DE/Ins/Per/Wichmann/Nightfall.html} program, presents the strongest variation strictly at the periastron passage. However, in the TIDES code, viscous effects that lead to a delay are taken into account. The most compact systems, model E1, the secondary of model E2, and model E3, remain nearly spherical during the entire orbital cycle (see Fig.\ref{fig:radiusE1E3}) because the mean radii are nearly constant and the highest relative difference between the maximum and minimum radii is less than $0.5$\%. This is not surprising because models E2, and E3 have the longest periods, hence the widest separation, and model E1 contains relatively compact main-sequence stars.
	
	\begin{table*}
		\caption{Parameters (from Antokhina et al. (\cite{Antokhina}), Rauw et al. (\cite{Rauw}), Gies et al. (\cite{Gies}), Sana et al. (\cite{Sanab}), Mayer et al. (\cite{Mayer}), and Sana et al. (\cite{Sanaa}) for models E1 to E5) of the eccentric binary systems. Inclinations in brackets stand for non-eclipsing systems.}
		\label{tab:PARAM}
		\centering
			\begin{tabular}{l c c c c c}
				\hline\hline 
				Parameters & Model E1 & Model E2 & Model E3 & Model E4 & Model E5\\
				\hline
				Period (day) & $6.08$ & $15.093$ & $9.6466$ & $5.604$ & $5.816$\\
				Eccentricity & $0.46$ & $0.234$ & $0.156$ & $0.259$ & $0.133$\\
				Argument of periastron of secondary (\degr) & $197.4$ & $202.5$ & $302$ & $284$ & $82$\\
				Mass ratio & $0.44$ & $0.54$ & $0.63$ & $0.74$ & $0.98$\\
				Inclination (\degr) & $(60)$ & $(32)$ & $(56)$ & $(60)$ & $67.2$\\
				Mass of primary ($M_{\sun}$) & $45$ & $68.5$ & $36.2$ & $24.5$ & $27.8$\\
				Mass of secondary ($M_{\sun}$) & $20$ & $37.3$ & $22.9$ & $18.2$ & $27.2$\\
				Primary polar temperature (K) & $49000$ & $39300$ & $40500$ & $32800$ & $34350$\\
				Secondary polar temperature (K) & $36500$ & $40100$ & $35000$ & $31200$ & $34000$\\
				Polar radius of primary ($R_{\sun}$) & $9.2$ & $24.0$ & $9.5$ & $10.3$ & $16.0$\\
				Polar radius of secondary ($R_{\sun}$) & $7.0$ & $10.0$ & $7.7$ & $7.9$ & $14.5$\\
				$v_1 sin(i)$ (km\,s$^{-1}$) & $135$ & $144$ & $102$ & $152$ & $135$\\
				$v_2 sin(i)$ (km\,s$^{-1}$) & $145$ & $75$ & $164$ & $133$ & $135$\\
				$\beta^\text{(a)}$ of primary & $0.64$ & $2.06$ & $1.80$ & $1.09$ & $0.81$\\
				$\beta^\text{(a)}$ of secondary & $0.91$ & $2.59$ & $3.57$ & $1.24$ & $0.89$\\
				\hline\hline
				TIDES code parameters &  &  &  &  & \\
				\hline
				Viscosity, $\nu$, of primary ($R_{\sun}^2$day$^{-1}$) & $0.05$ & $0.045$ & $0.028$ & $0.05$ & $0.05$\\
				Viscosity, $\nu$, of secondary ($R_{\sun}^2$day$^{-1}$) & $0.05$ & $0.01$ & $0.028$ & $0.02$ & $0.02$\\
				Layer depth & $0.01$ & $0.1$ & $0.07$ & $0.02$ & $0.1$\\
				Polytropic index of primary & $1.5$ & $3$ & $1.5$ & $3$ & $3$\\
				Polytropic index of secondary & $1.5$ & $1.5$ & $1.5$ & $1.5$ & $3$\\
				Number of azimuthal $\times$ latitudinal partitions  & $500\times20$ & $500\times20$ & $500\times20$ & $500\times20$ & $500\times20$\\						
				\hline
			\end{tabular}
			\tablefoot{$^\text{(a)}$The $\beta$ parameter measures the asynchronicity of the star at periastron and is defined by $\beta = 0.02\frac{Pv_{rot}}{R}\times \frac{(1-e)^{3/2}}{(1+e)^{1/2}}$, where $v_{rot}$ is the equatorial rotation velocity, $R$ is the equilibrium radius, and $e$ is the eccentricity.}
	\end{table*}

	\subsection{Spectral classification}
	First, we applied the quantitative Conti--Mathys criterion (Conti \& Alschuler \cite{Conti}, Mathys \cite{Mathys(1)}, \cite{Mathys(2)}). The spectral types that we found for our simulated spectra either agree well with the observations or are hotter than the observations. The hotter stars present a weak He\,I $\lambda$\,4143. This makes measuring the spectra of this system difficult and uncertain. The luminosity classes are, as in Paper I, overestimated for models E1, E3, and E5. Model E2 and the primary of model E1 are too hot for the Conti--Mathys luminosity criterion. The luminosity classes agree well for model E4. The problem of simulating singlet transition of He\,I reported by Najarro et al. (\cite{Najarro}) and discussed in Paper I is probably responsible for the general luminosity classification problem. To refine our classification, we again used the Walborn \& Fitzpatrick (\cite{Walborn}) atlas. This atlas was used with the individual spectra, which for real observational data can only be assessed through spectral disentangling. The atlas is more qualitative than the Conti--Mathys criterion but relies on a large number of lines. We can now achieve a better agreement with the classification of the real binary systems (see Table \ref{tab:SpecClass}).

		\begin{table*}[ht!]
			\caption{Spectral classification using the Conti--Mathys criterion and the Walborn \& Fitzpatrick atlas.}
			\label{tab:SpecClass}
			\centering
				\begin{tabular}{l c c c c}
					\hline\hline
					Stars & Combined spectra & Individual spectra & W\&F atlas & Observational analysis\\
					\hline 
					Model E1$^{\text{1}}$ & O3 & O3 & O3-4 (IV)-V & O3V\\
					Model E1$^{\text{2}}$ & O7.5I & O7I & O8-8.5 III-(V) & O8V\\
					Model E2$^{\text{1}}$ & O5.5 & O4 & O5-6 (I)-III & O5.5I\\
					Model E2$^{\text{2}}$ & O4 & O4 & O7-8 III-(V) & O7V\\
					Model E3$^{\text{1}}$ & O6I & O5.5I & O6.5-7 III-(V) & O6.5V\\
					Model E3$^{\text{2}}$ & O6.5I & O7I & O8.5 V & O8.5V\\
					Model E4$^{\text{1}}$ & O8I & O8.5III & O9-9.5 (III)-V & O9III\\
					Model E4$^{\text{2}}$ & O9III & O9.5V & O9.5-9.7 III-(V)& O9.7V\\
					Model E5$^{\text{1}}$ & O6.5I & O7I & O7-8 III-(V) & O7III\\
					Model E5$^{\text{2}}$ & O7I & O7I & O7.5-8.5 (III)-V & O7.5III\\
					\hline
				\end{tabular}
				\tablefoot{$^1$Primary. $^2$Secondary.}
		\end{table*}
		
	\subsection{Struve-Sahade effect }
	In our sample of binaries, two are known to present this effect: HD\,93403 and HD\,152248. They inspired our models E2 and E5. In our analysis, we detected variations of the line strength in models E1, E2, E4, and E5. However, we observe an S-S effect in the sense of Linder et al. (\cite{Linder}) only for some lines in model E2.
		
		For all models, we investigated eight ``well-chosen'' lines in the spectra at 20 phases (the phase zero corresponds to the periastron passage). The lines were chosen following three criteria: first, we selected the lines for which an S-S effect is reported in the literature. The second criterion is based on a visual detection of variations in the line profile in the synthetic spectra, and finally, the third criterion is based on the fact that some lines present the S-S effect more often than others. As in Paper I, we draw attention to possible blends with nearby lines that can modify the strength and lead to misinterpretations. 
		
\begin{figure}
		  \centering
			\includegraphics[width=6cm]{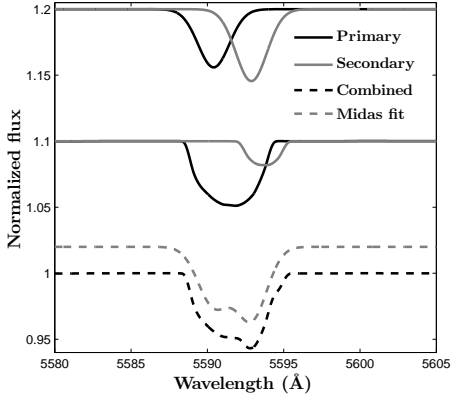}
			\caption{Combination of the O\,III $\lambda 5592$ line of the primary and the secondary for model E2 at phase $0.25$. The deblend/line routine will not fit the line of the secondary properly. \textit{Top:} The resulting lines of the two Gaussian fits. \textit{Middle:} Individual lines of the primary and secondary computed with our model and normalized to the continuum of the entire system. \textit{Bottom:} Spectrum of the system and corresponding fit of MIDAS.}
			\label{fig:linescombE2}
		\end{figure}

\begin{figure}[ht!]
		  \resizebox{\hsize}{!}{\includegraphics{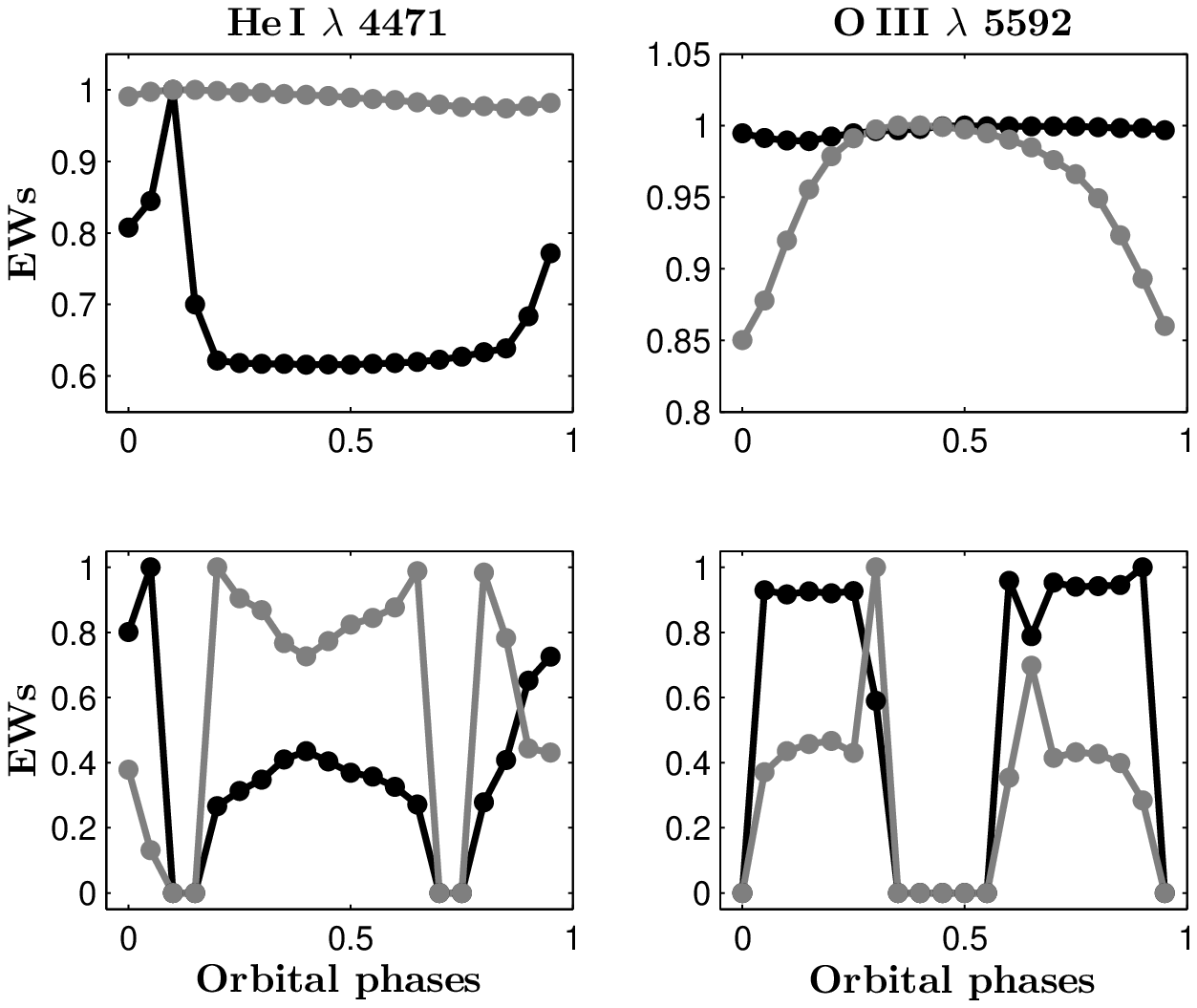}}
			\caption{Example of EW variations for the He\,I $\lambda$ 4471 line for model E2 and for the O\,III $\lambda$ 5592 line for model E4. The EWs have been normalized to the highest value. \textit{Black:} Primary. \textit{Grey:} Secondary. \textit{Top}: Variations measured on individual spectra (EW$_{\text{p, max, HeI}}=0.17$\AA, EW$_{\text{p, max, OIII}}=0.18$\AA, EW$_{\text{s, max, HeI}}=0.34$\AA, EW$_{\text{s, max, OIII}}=0.11$\AA). \textit{Bottom}: Variations measured on combined spectra (EW$_{\text{p, max, HeI}}=0.17$\AA, EW$_{\text{p, max, OIII}}=0.14$\AA, EW$_{\text{s, max, HeI}}=0.10$\AA, EW$_{\text{s, max, OIII}}=0.08$\AA).}
			\label{fig:EWmodelE2E4}
		\end{figure}

\begin{figure}[ht!]
		  \resizebox{\hsize}{!}{\includegraphics{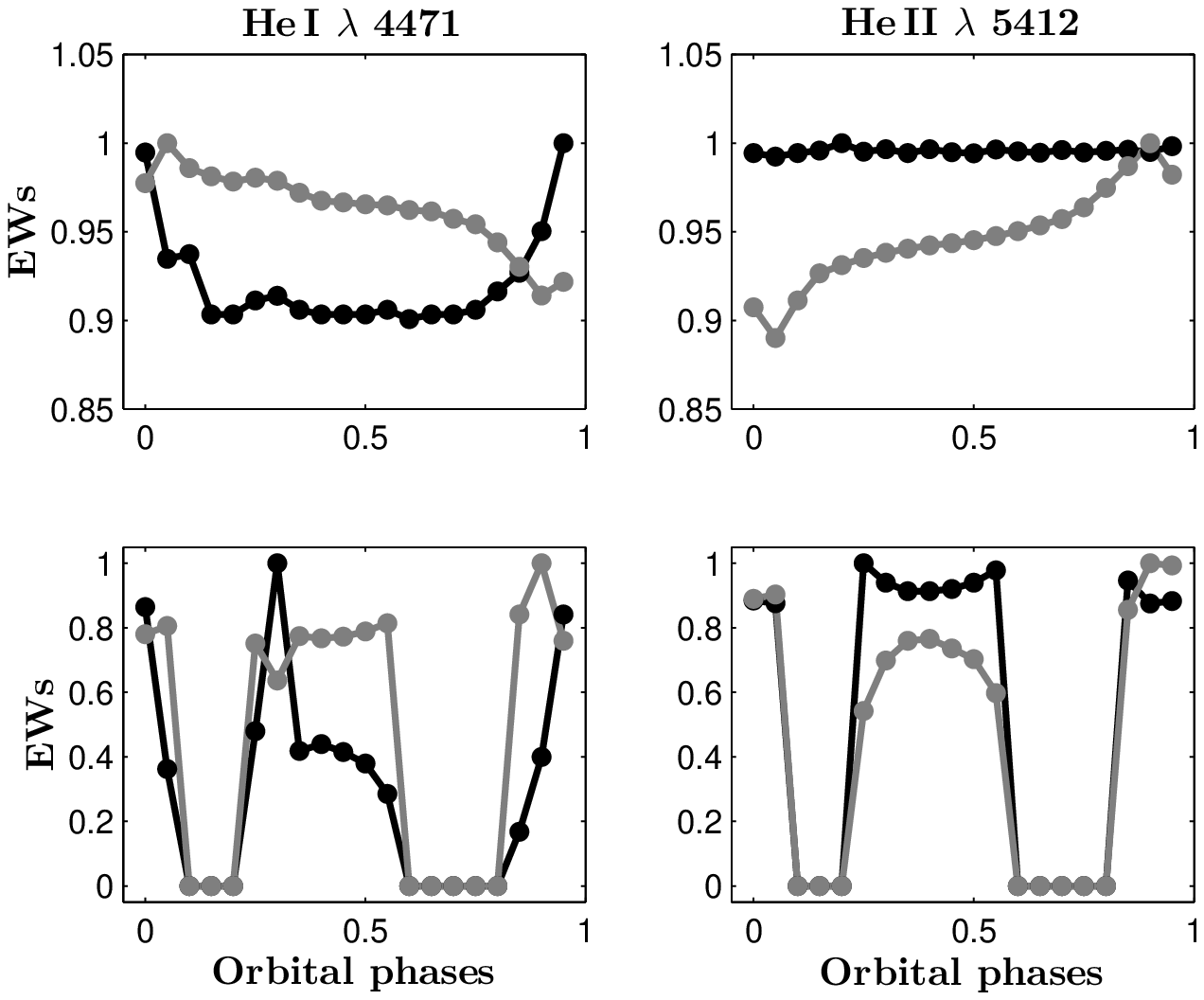}}
			\caption{Example of EW variations for the He\,I $\lambda$ 4471 and He\,II $\lambda$ 5412 lines for model E1. The EWs have been normalized to the highest value. \textit{Black:} Primary. \textit{Grey:} Secondary. \textit{Top}: Variations measured on individual spectra (EW$_{\text{p, max, HeI}}=0.04$\AA, EW$_{\text{p, max, HeII}}=1.07$\AA, EW$_{\text{s, max, HeI}}=0.59$\AA, EW$_{\text{s, max, HeII}}=0.98$\AA). \textit{Bottom}: Variations measured on combined spectra (EW$_{\text{p, max, HeI}}=0.05$\AA, EW$_{\text{p, max, HeII}}=0.80$\AA, EW$_{\text{s, max, HeI}}=0.21$\AA, EW$_{\text{s, max, HeII}}=0.29$\AA).}
			\label{fig:EWmodelE1}
		\end{figure}

	We followed the same procedure as in Paper I and measured the EWs in the simulated combined spectra of the binary at different phases as well as in the simulated spectra of the individual components of the binary at the same phases. This approach allows us to compare the line strengths that are deblended from the combined spectrum with the actual individual spectra, which is not possible with the real observational data.
	We found variations in the EWs in many of the individual spectra as well as in the combined spectra. However, the agreement between the results of the deblend/line routine and the actual individual spectra is mostly poor. We present in Figs. \ref{fig:EWmodelE2E4} and \ref{fig:EWmodelE1} examples of typical variations observed in our analysis. The (inverse) U-pattern (Fig. \ref{fig:EWmodelE2E4} \textit{right, grey} or Fig.\ref{fig:EWmodelE1} \textit{left, black}) is often observed and is directly related to the orbital motion. Constant EWs or irregular variations are also often observed. A constant EW is mostly observed in long-period systems. We also encountered other types of variation patterns such as those presented in Fig.\ref{fig:EWmodelE1} (\textit{right, grey}) during our analysis. The amplitude of the EW variations depends on the line and the system. The variations observed on individual spectra can be explained by the physical variations of the temperature and gravity at the stellar surface during the orbital cycle (we give an example of the interpretation of the variation in subsection \ref{sssec:modelE1}). 
	 The explanation of the S-S effect that we suggested in Paper I seems to be reinforced by these new analyses. The S-S effect seems to be due to the combination of the spectra in which the lines have a non-Gaussian/non-Lorentzian and even asymmetric profile. Under these conditions, the deblend/line routine introduces systematic errors that mimic the S-S effect. This is illustrated in Fig. \ref{fig:linescombE2}, which shows that the deblending routine erroneously selects the bump that is created by the superposition of the two lines as the secondary star's absorption line. The secondary's line appears as a weak contribution on the red wing of the primary star's absorption. However, a fit with two Gaussian profiles will consider the bump as the core of the secondary line and thereby overestimate its strength.
	
		\subsubsection{Model E1} \label{sssec:modelE1}
		In this model, we mainly observed variations illustrated in Fig.\ref{fig:EWmodelE1}. These variations can be explained by the orbital motion of the stars and the variation of the visible part of the stars. 
		At periastron passage, we see the rear side of the secondary. Because the stars are close, they are quite deformed and therefore the visible part of the secondary is cool, which leads to a reinforcement of the He\,I lines. For the primary the hemisphere facing the companion is visible at this phase. However, because the primary is deformed, and owing to the large temperature difference between the two stars, the reflection process does not heat the primary significantly. Thus it is also the coldest part of the primary that is visible at this phase. When the binary separation increases, the stars become nearly spherical. The temperature of the primary becomes nearly constant and the EW of its lines do not change. However, for the secondary, we begin to see the front part of the star. Whilst the reflection is not very effective for heating the primary, it is very effective for the secondary and accordingly, when we see the front part of the latter, the temperature increases. The increase is stronger because the separation between the stars decreases during the second half of the orbital cycle.
		 
		 The EWs measured on combined spectra (see Fig.\ref{fig:EWmodelE1}, \textit{Bottom}) are not consistent with those measured on the individual spectra (except for the He\,I $\lambda$\,4143 line of the secondary). As mentioned above, this inconsistency is due to the deblending routine, which does not properly separate the lines of the primary and secondary stars. For example, the blue-shifted component may have a larger EW than when it is red-shifted. This inconsistency is reduced when the lines are measured at orbital phases at which they are clearly separated.
		 
	  \subsubsection{Model E5}
	For this last model, we investigated the He\,I $\lambda\lambda$\,4026, 4143, 4471, 4713, 5016, Si\,IV $\lambda$\,4089, He\,II $\lambda$\,4200, 4542, and O\,III $\lambda$\,5592 lines. The EWs of the He\,I and Si\,IV lines in the primary display a U-pattern with a maximum at phase $0.0-0.05$ and a minimum at phase $0.45-0.55$. The EWs of the secondary display a symmetric variation with respect to the primary, however, the amplitude of the variation is smaller. The O\,III line displays the inverse behaviour compared to the previous lines. Finally, the He\,II lines in the primary display two maxima at phases $0.3$ and $0.7$ with an intermediate minimum at phase $0.5$ that give us an M-pattern variation. The secondary presents a low-amplitude U-pattern variation.
	 
		\begin{figure}
		  \resizebox{\hsize}{!}{\includegraphics{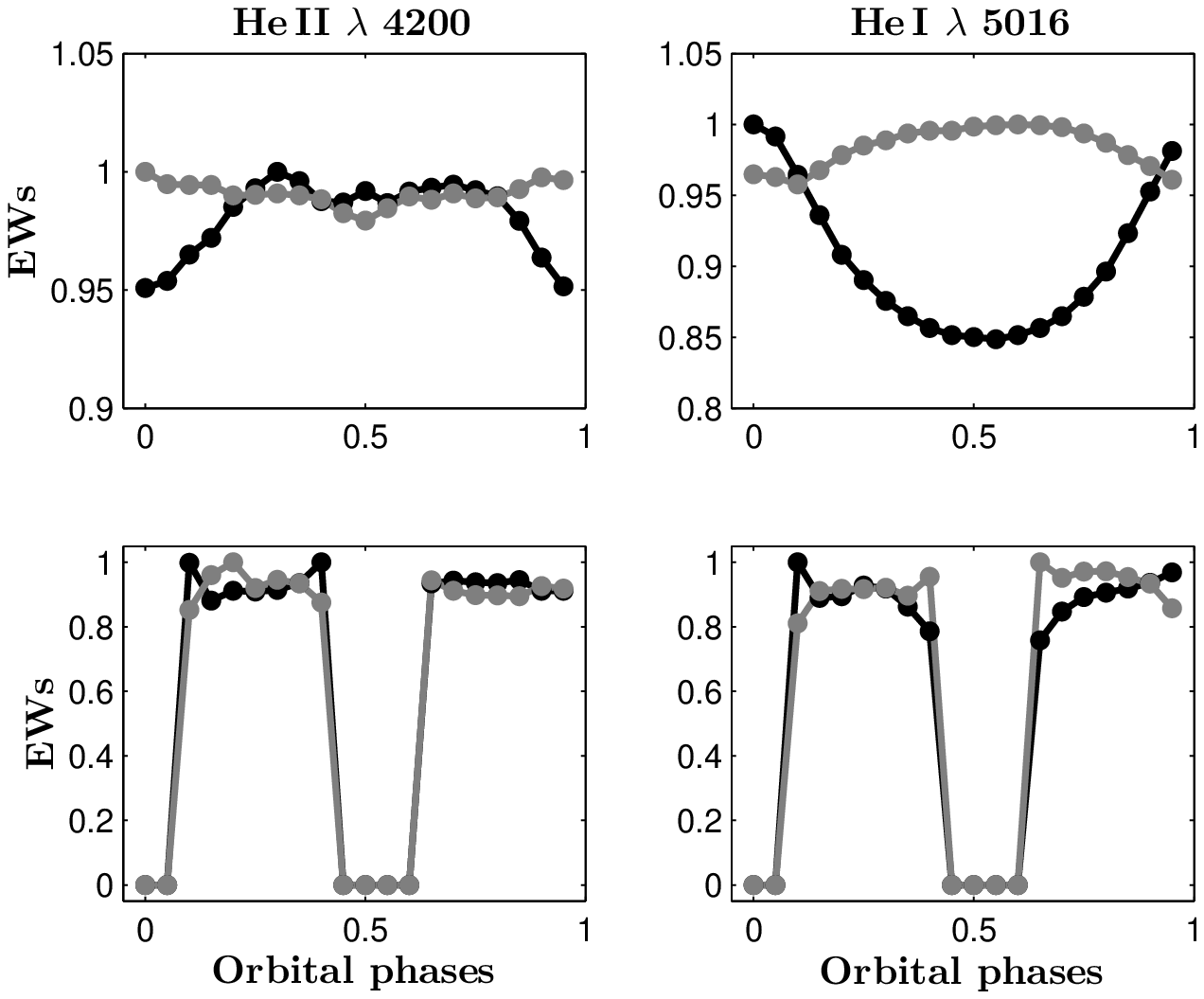}}
			\caption{Example of EW variations for He\,I $\lambda$ 5016 and He\,II $\lambda$ 4200 for model E5. The EWs have been normalized to the highest value. \textit{Top}: Variations measured on individual spectra (EW$_{\text{p, max, HeI}}=0.20$\AA, EW$_{\text{p, max, HeII}}=0.48$\AA, EW$_{\text{s, max, HeI}}=0.21$\AA, EW$_{\text{s, max, HeII}}=0.48$\AA. \textit{Bottom}: Variations measured on combined spectra (EW$_{\text{p, max, HeI}}=0.12$\AA, EW$_{\text{p, max, HeII}}=0.27$\AA, EW$_{\text{s, max, HeI}}=0.11$\AA, EW$_{\text{s, max, HeII}}=0.24$\AA).}
			\label{fig:EWmodelE5}
		\end{figure}
	 
	The EWs measured on the combined spectra are nearly constant for the He\,I $\lambda\lambda$\,4026, 4471, 5016, He\,II $\lambda$\,4200, 4542, and O\,III $\lambda$\,5592 lines for the primary and secondary stars. The Si\,IV line displays irregular variations but a mean value of the primary higher than the secondary for phases before $0.5$ and a mean value of the secondary higher for the second part of the orbital cycle. The He\,I $\lambda$\,4713 line displays constant EWs for phases before and after $0.5$ but the value is different before and after phase $0.5$. Before phase $0.5$ the primary has a lower EW than the secondary and after phase $0.5$ the primary has a larger EW than the secondary. Finally, the variations observed for the He\,I $\lambda$\,4143 line are consistent with the variations measured on individual spectra. All variations measured in our model underestimate the observed S-S effect.
	
	In conclusion, our models display, in general, some variations of EWs during the orbital cycle. The variations on the spectra of individual components are caused by the orbital motion and the modulation of the visible part of the star. However, the variations that we measure on combined spectra and the visual variation of the relative strength of the line of the primary and secondary stars are not strong enough to completely explain the S-S effect as observed by Sana et al. (\cite{Sanaa}).
		
	\subsection{Synthetic broad-band light curves and radial velocity curve}
\begin{figure}
		  \resizebox{\hsize}{35mm}{\includegraphics{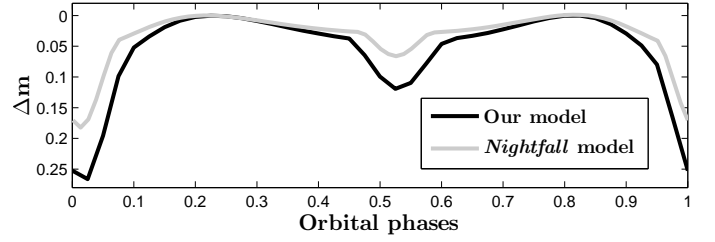}}
			\caption{Synthetic light curves of model E5 in the range $\lambda\lambda$ 3500, 7100. The light curves have been computed with the CoMBiSpeC + TIDES models (\textit{black}) and with the Nightfall model (\textit{grey}).}
			\label{fig:lightcurvesE}
		\end{figure}

	The synthetic light curves were computed in the wavelength range $3800-7100$ \AA. We compared our light curves to those obtained with \textit{Nightfall}. \textit{Nightfall} provides synthetic light curves based on an instantaneous Roche potential and thus does not take any viscous stress into account. We point out here the similarity and difference of the two models. For models E1 and E4, our light curves present qualitatively the same characteristics as \textit{Nightfall}. The TIDES code leads to more compact objects than the instantaneous Roche potential. Therefore, because the stars are more compact, all other parameters being equal, the gravity darkening is less strong and the reflection effect is more effective and consequently the stars are hotter. Near periastron, the difference between the two models is stronger because our model predicts a weak variation of $0.02$ mag and \textit{Nightfall} predicts no variation. The very low inclination (60$\degr$) explains this small change of magnitude. Because the stars are less deformed (than in the Nightfall model), the reflection effect is more effective near periastron and accordingly, it also increases the temperature of the stars which enhances the difference between the two models. Some other differences arise because of the time delay induced by viscous stress. For model E1, the agreement between the two models and the light curve observed for HD\,93205 by Antokhina et al. (\cite{Antokhina}) is poor, however. Models E2 and E3 give very different results with our code and \textit{Nightfall}. These two systems have a longer period and are more compact in our model than in \textit{Nightfall}. The discrepancies probably come from the different radii of the stars. 
	Finally, the light curve of model E5 displays the two eclipses in both models but in our model, the eclipses are deeper than evaluated with \textit{Nightfall} (see Fig.\ref{fig:lightcurvesE}). Again the difference can be explained by the smaller radius of the star in our model.	Some other small differences can be easily explained by the viscous effects. The agreement between our model and the observations of Mayer et al. (\cite{Mayer}) is rather good, though we slightly overestimate the depth of the eclipse of the primary.
	
	The mean brightness ratios in the wavelength range $3800-7100$ \AA\ are similar to the literature values. Finally, we determined the amplitude of the radial velocity curves and found a good agreement with the observations (see Table \ref{tab:K}) for all simulated systems.

		\begin{table}[ht!]
			\caption{Semi-amplitude of radial velocity curves for eccentric models and comparison with observations (in km\,s$^{-1}$)}
			\label{tab:K}
			\centering
				\begin{tabular}{l c c c c}
					\hline\hline
					Stars & K$_1$ & K$_2$ & K$_{1\text{, obs}}$ & K$_{2\text{, obs}}$\\
					\hline 
					Model E1 & $136.3$ & $308.8$ & $132.6$ & $313.6$\\
					Model E2 & $77.3$ & $142.8$ & $79.3$ & $139.0$\\
					Model E3 & $125.3$ & $197.5$ & $117.0$ & $211.0$\\
					Model E4 & $157.6$ & $213.9$ & $162.4$ & $213.9$\\
					Model E5 & $209.8$ & $208.4$ & $206.9$ & $211.7$\\
					\hline
				\end{tabular}
		\end{table}

\section{Summary and perspectives} \label{sec:5}

 We have presented improvements of our mathematical model that allow us to compute the physical properties on the surface of massive stars in binary systems that contain main-sequence O(B) stars. The first improvement is the inclusion of the radiation pressure effect on the shape of the stars. The second improvement is the use of the TIDES code to compute the shape of the stars in eccentric and/or asynchronous systems. In both cases, we took into account various effects such as gravity darkening, reflection, and limb-darkening which allowed us to compute the temperature distribution at the stellar surface. Then we used the TLUSTY OSTAR2002 and BSTAR2006 grids (Lanz \& Hubeny \cite{Lanza}, \cite{Lanzb}) to compute the spectra of each star of the system as a function of orbital phase.
We showed that in a number of cases, the radiation pressure does not have a strong impact on the shape of the stars and therefore on the spectra for the models studied in Paper I. This implies that the conclusions of Paper I for the S-S effect remain valid. Our results showed that the radiation pressure has a weak impact on the shape of highly deformed stars of (over)contact binary systems like model 3. However, in model 3, the small changes in the shape of the stars have resolved our previous problem of the surface temperature distribution that we failed to reproduce with our first version of the algorithm. We also studied the impact of the gravity-darkening parameter variation on the spectra of model 3. Our results indicate that, given the assumptions made for the computation, there is no need to use a more complex treatment of the gravity darkening than the classical von Zeipel theorem. The model 4 clearly showed that radiation pressure can have a strong impact on the shape of the stars. Many of the spectral lines in this model display phase-locked profile and/or strength variations. However, these variations are not sufficient to reproduce the S-S effect observed by Bonanos (\cite{Bonanos}).

The second part of this paper investigated the eccentric systems. We studied many lines and detected phase-locked profiles and/or strength variations in many cases in the individual spectra of both components. These variations are caused by the change of orientations of the stars as a function of orbital phase. They thus reflect the non-uniform temperature distribution across the stellar surface. The variations measured on the combined spectra of the systems, however, often disagree with the measurements made on individual spectra. Our results also showed that the deblending routines that fit two Gaussian or Lorentzian profiles often fail to measure the line EWs properly and lead to incorrect interpretations. The variations measured in models E2 and E5 underestimate the S-S effect observed in HD\,93403 and HD\,152248, which in turn might be caused by our not predict intrinsic asymmetric line profiles in these cases. In Paper I, this asymmetry was our explanation for the S-S effect: the sum of asymmetric intrinsic line profiles of individual stars are not properly deblended by Gaussian/Lorentzian profile fits. The absence of this asymmetry in eccentric systems is, however, not easy to explain and could be linked to weaker interactions between the stars in eccentric binaries than in close circular ones. Therefore, we cannot generalize the interpretation of the S-S effect of Paper I to the eccentric systems studied here. However, this highlights that analyses of binary system spectra are very difficult and some observational effects could be generated by analysing techniques and not by physical processes.

For clearly detached systems, spectral disentangling (Hadrava \cite{Hadrava}, Gonz\'alez \& Levato \cite{Gonzalez}, Simon \& Sturm \cite{Simon}) helps to overcome the difficulties of the deblending routines. However, for systems with a strong temperature gradient at their surface (e.g. HD\,100213, Linder et al. \cite{Linder}), the technique fails because different lines have different radial velocity amplitudes. In eclipsing binaries, disentangling cannot be used for the phases near the eclipses. This technique provides mean spectra of both components, and even if we can measure variations with respect to this mean, we cannot compute the spectra at each orbital phase.

A future step could be to introduce cross-talk which is surely substantial in eccentric and asynchronous systems. Finally, we plan to include a wind interaction zone between the stars which could contribute to the heating of the stellar surface in two ways, either by backscattering of the photospheric photons, or by irradiation of X-ray photons emitted by the shock-heated plasma in the wind interaction zone. This requires, however, more sophisticated atmosphere models than what we have employed so far.

\begin{acknowledgements}
MP and GR acknowledge support through the XMM/INTEGRAL PRODEX contract (Belspo), from the Fonds de Recherche Scientifique (FRS/FNRS), as well as by the Communaut\'e Fran\c caise de Belgique - Action de recherche concert\'ee - Acad\'emie Wallonie - Europe. GK and EM acknowledge support from UNAM/PAPIIT 107711.
\end{acknowledgements}


\end{document}